\documentclass[runningheads]{llncs}
\usepackage{graphicx}
\usepackage{booktabs}
\usepackage[table,xcdraw]{xcolor}
\usepackage{multirow}
\usepackage{tabularx}
\usepackage{hyperref}
\usepackage{amsmath}
\usepackage{array}
\usepackage{xr}
\usepackage{siunitx}
\usepackage{parskip}

\begin{document}
\title{Physics-Informed Neural Networks can accurately model cardiac electrophysiology in 3D geometries and fibrillatory conditions}
\titlerunning{PINNs for Cardiac Electrophysiology}
%
\author{Ching-En Chiu\inst{1,2} \and Aditi Roy\inst{3} \and Sarah Cechnicka\inst{4} \and Ashvin Gupta\inst{5} \and Arieh Levy Pinto\inst{5} \and Christoforos Galazis\inst{4} \and Kim Christensen\inst{6} \and Danilo Mandic\inst{1} \and Marta Varela\inst{2}}
 \authorrunning{Chiu \textit{et al}.}
%

\institute{Department of Electrical and Electronic Engineering, Imperial College London, UK \and 
National Heart \& Lung Institute, Imperial College London, UK \and
Department of Biomedical Engineering, Kings College London, UK \and
Department of Computing, Imperial College London, UK \and
Department of Bioengineering, Imperial College London, UK \and
Department of Physics, Imperial College London, UK 
\\
\email{marta.varela@imperial.ac.uk}}
\maketitle              
\begin{abstract}
Physics-Informed Neural Networks (PINNs) are fast becoming an important tool to solve differential equations rapidly and accurately, and to identify the systems parameters that best agree with a given set of measurements. PINNs have been used for cardiac electrophysiology (EP), but only in simple 1D and 2D geometries and for sinus rhythm or single rotor dynamics. Here, we demonstrate how PINNs can be used to accurately reconstruct the propagation of cardiac action potential in more complex geometries and dynamical regimes. These include 3D spherical geometries and spiral break-up conditions that model cardiac fibrillation, with a mean RMSE $< 5.1\times 10^{-2}$ overall. 

We also demonstrate that PINNs can be used to reliably parameterise cardiac EP models with some biological detail. We estimate the diffusion coefficient and parameters related to ion channel conductances in the Fenton-Karma model in a 2D setup, achieving a mean relative error of $-0.09\pm 0.33$. Our results are an important step towards the deployment of PINNs to realistic cardiac geometries and arrhythmic conditions.   


\end{abstract}

\section{Introduction}
Physics-Informed Neural Networks (PINNs) are a machine learning method that integrates data-driven learning with knowledge of the physical equations describing a system \cite{Raissi2019}. This domain knowledge is explicitly incorporated in the loss function of the neural network (NN). This ensures that PINNs' inferences are consistent with the physical understanding of a system and enables learning with only a small fraction of the data that conventional NNs require.\\

In cardiac electrophysiology (EP), PINNs have been used with eikonal models to predict arrival times of action potential in the left atrium. Sahli Costabal \textit{et~al.} initially estimated high-resolution arrival time maps using this approach \cite{SahliCostabal2020Physics-InformedMapping}. Herrera \textit{et~al.} then built on it to estimate atrial fibre orientations 
\cite{RuizHerrera2022Physics-informedMaps}. Of most relevance to our study is the research of Herrero Martin \textit{et~al.}, which used PINNs with the monodomain equation on sparse maps of transmembrane potential to estimate EP parameters (such as the isotropic diffusion coefficient or surrogates of the action potential duration) \cite{HerreroMartin2022EP-PINNs:Networks}. This study was limited to 1D and 2D geometries and relied on the Aliev-Panfilov model, a simple two-variable model. In another work, PINNs were coupled with a more biologically detailed EP model, the Fenton-Karma model~\cite{fenton1998vortex}, to characterise the effects of anti-arrhythmic drugs. PINNs successfully estimated the effect of drugs on EP parameters related to the conductance of different ionic channels~\cite{chiu2024characterisation}. This work demonstrated PINNs' capability of working with experimental \textit{in vitro} data, but was nonetheless limited to a simple 1D cable geometry.

To provide clinically useful characterisation of cardiac EP properties, PINNs will need to be deployed in 3D, using more biologically detailed EP models, and in both sinus rhythm and fibrillatory conditions. This will improve our understanding of the mechanisms of arrhythmias and help design personalised treatments that target regions with abnormal EP parameters. This is the gap this study aims to address. 

\subsubsection{Aims}
We aim to use PINNs to predict the spatial-temporal propagation of cardiac action potentials:
\begin{enumerate}
    \item Using the two-variable Aliev-Panfilov (AP)  model~\cite{aliev1996simple},
   \begin{enumerate}
       \item in 3D spherical geometry, for a centrifugal wave (modelling sinus rhythm).
       \item in 3D spherical geometry, for a single spiral wave (modelling tachycardia).
       \item in 2D rectangular geometry, for spiral wave break-up (modelling fibrillatory conditions).
   \end{enumerate} 
    \item Using the three-variable Fenton-Karma (FK) model~\cite{fenton1998vortex}, 
    \begin{enumerate}
        \item in 2D rectangular geometry, for a planar wave (sinus rhythm).
        \item in 2D rectangular geometries, for a single spiral wave (tachycardia).
    \end{enumerate}
     For the FK model, we additionally use PINNs in inverse mode to simultaneously estimate global EP parameters, as the FK model can provide insights into more detailed EP properties. These include the diffusion coefficient, $D$, and parameters representing the conductance of the sodium, calcium and potassium channels. We estimate each parameter one at a time. In the following aims:
     \item We use PINNs to predict the action potential propagation with one unknown EP parameter, and estimate the parameter simultaneously:
     \begin{enumerate}
         \item in 2D rectangular geometry, for a planar wave (sinus rhythm).
        \item in 2D rectangular geometries, for a single spiral wave (tachycardia).
        \end{enumerate}
    \end{enumerate}


Next, we briefly introduce the AP and FK cell models used to model cardiac action potentials in the monodomain formulation. The full equations of the two models can be found in the Supplementary Materials.

\subsubsection{Aliev-Panfilov model} The model consists of one partial differential equation (PDE) and one ordinary differential equation (ODE), coupled together, describing fast and slow processes. The two equations describe the evolution of two variables: transmembrane potential $V$ and a non-observable recovery variable $W$, related to the restitution, which enforces refractoriness properties. The variables are dimensionless: $V$ is scaled to the $[0, 1]$ interval (AU). As in Ref~\cite{aliev1996simple},
we used a temporal unit (TU) that corresponds to approximately \num{13}\unit{\milli\second}. 



\subsubsection{Fenton-Karma model}
The FK model is a more complex EP model consisting of one PDE and two ODEs, all coupled together. It has explicit formulations of three transmembrane currents: a fast inward $J_\text{fi}$, slow outward $J_\text{so}$, and slow inward $J_\text{si}$. These are analogous to the Na$^+$, K$^+$, and Ca$^{2+}$ currents, respectively. 

The three DEs describe the evolution of three variables: $u$, $v$, and $w$. Variable $u$ is the dimensionless membrane potential scaled to $[0,1]$ (corresponding to $V$ in the AP model). Time $t$ is in \unit{\milli\second}. $v$ and $w$ are latent gate variables for $J_\text{fi}$ and $J_\text{si}$, respectively. There are various parameters in the model equations: $\tau_d$, $\tau_o$, $\tau_r$, and $\tau_{si}$, which are approximately inversely related to the conductance of different ion channels.

For both models, we adopt the no-flux Neumann boundary condition: $\frac{\partial V}{\partial \vec{n}} = 0$, which enforces no leakage of $V$ outside of the domain (equivalently $u$ in the FK model). We assume homogeneous and isotropic diffusion of the membrane potential throughout, and the diffusion coefficient $D$ is therefore a scalar. 

\section{Methods}
All code used in this study is available at: \url{https://github.com/annien094/2D-3D-EP-PINNs}. 

\subsection{Ground Truth Data Generation}

Ground truth \textit{in silico} maps of membrane potential $V$ (or equivalently $u$ in the FK model) in various geometries and dynamics are generated using in-house code written in Matlab. These generated data are used for the training and testing of the PINNs. 20\% of the ground truth data are uniformly selected across time at each location to train the PINN solver. The remaining 80\% are used to test the trained PINNs. The values of parameters used are detailed in Supplementary Table 1 \& 2.

\subsubsection{3D Spherical Surface Geometry} The 3D geometry is a homogeneous spherical shell, with an inner radius of \qty{10}{\cm} and outer radius of \qty{12}{\cm}. We use the finite element method (FEM) in the Matlab PDE Toolbox to solve the AP model on a tetrahedral mesh (7670 elements, 13669 nodes, average edge length \qty{1.61}{\cm}). An explicit forward solver ($\text{d}t=0.5$ TU) is used for the temporal domain. 

We initialise electrical activity in two different ways. In the first case modelling sinus rhythm, the $V$ of a small cuboid area (\qty{1}{\cm}-side) around the north pole is raised at $0.8$ AU at $t=0$. This generates a wave that travels from the north pole down the shell with a circular wavefront. We will refer this type of wave as  “centrifugal wave". In the second case modelling tachycardia, we use a cross-field protocol to generate a spiral wave: first generate a centrifugal wave; after some time, reset $V$ for half of the sphere to zero to generate sustained spiral waves. 
\subsubsection{2D Rectangular Geometry} The 2D geometry is a square with \qty{10}{\cm} sides. We use central finite differences coupled with a forward Euler solver to solve the EP models. For the AP model, we use $\text{d} x=\qty{0.1}{\mm}$ and $\text{d}t=0.05$~TU as the spatial and temporal steps; for the FK model, we use $\text{d}x=\qty{0.3}{\mm}$ and $\text{d}t=\qty{0.01}{\ms}$. 

Planar waves are initiated by applying a supra-threshold external stimulus for $1$ TU from one side of the rectangle. Spiral waves are created using the cross-field protocol. In the AP model simulations (Aim 1c), we observed that the generated spiral waves broke up spontaneously for the model parameters used.


\subsection{PINNs Setup}
We implement PINNs with the DeepXDE library~\cite{lu2021deepxde} using a TensorFlow back-end. We use a fully connected NN trained with the Adam optimiser for \num{150000} epochs, followed by the L-BFGS-B optimiser to facilitate convergence. Glorot initialisation and a tanh activation function are employed. 

\begin{figure}[htp]
\centering
\includegraphics[width=0.8\textwidth]{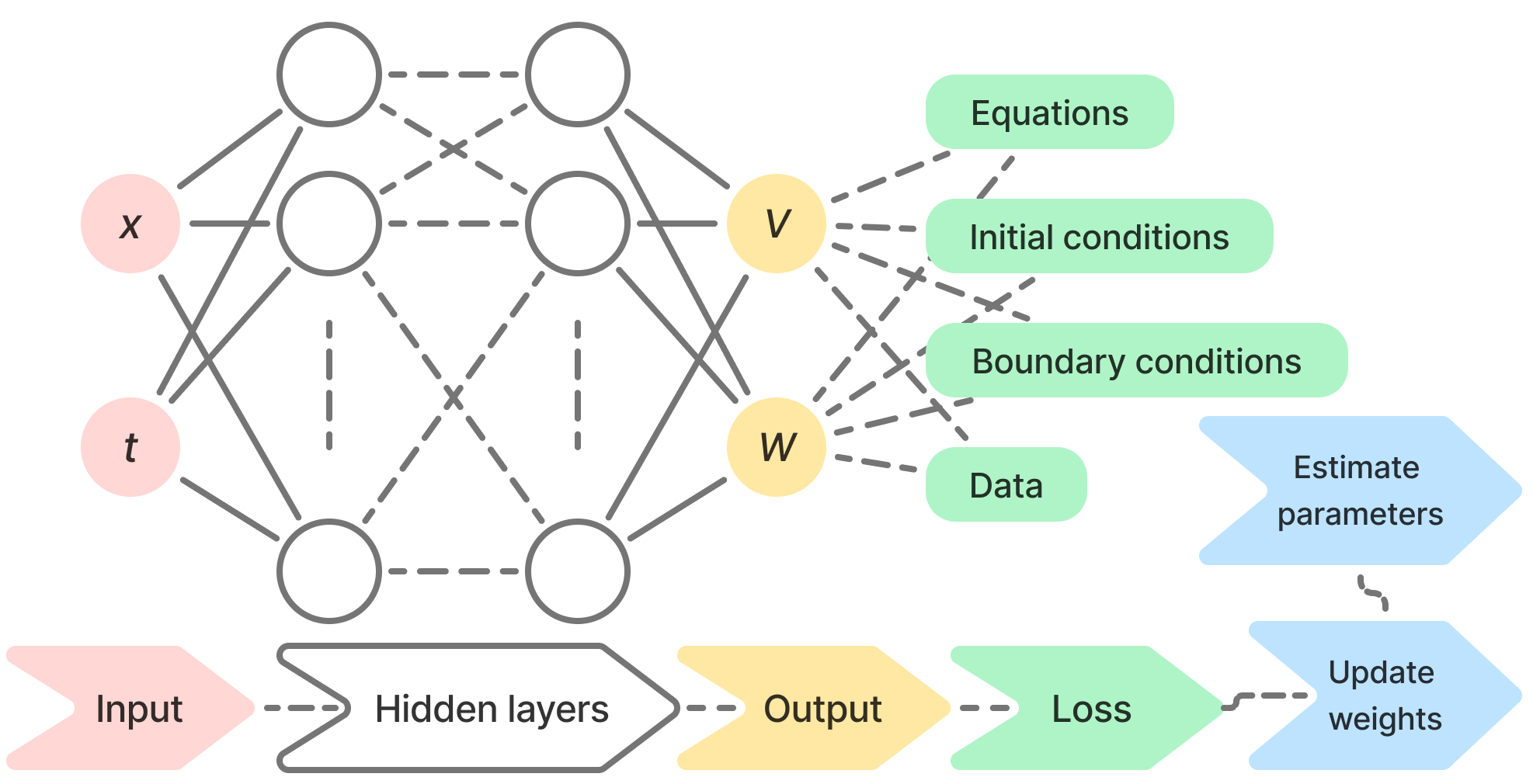}
\caption{The PINN architecture, using the AP model as an example for the output variables. The network takes the spatial and temporal location $x$ and $t$ as inputs, and outputs $V$ and $W$. For the FK model, output variables are $u$, $v$, and $w$ instead. PINN optimises its weights based on four different loss terms shown in \eqref{eq:loss}.  In inverse mode (Aims 3a \& 3b), PINN also estimates one unknown parameter.}
\label{fig:PINN arch}
\end{figure}

A schematic diagram of PINN's architecture is shown in Fig.~\ref{fig:PINN arch}. The inputs are space, $x$, and time, $t$, position coordinates. For the AP model, there are two output variables:
$V$ and $W$; for the FK model there are three: $u$, $v$, and $w$. In the 2D spiral break-up regime (Aim 1c), the network size is 5 layers of 60 neurons each, as we found that a larger network is needed for the more complex dynamics. For the rest of the scenarios, the network size is 4 layers of 32 neurons each. We apply a transform to the input layer: $t$ $\rightarrow$ a vector of $\sin (kt)$ ($k=1,2,...$, up to the number of hidden layers)~\cite{lu2021deepxde}. This helps capture the periodicity of the solutions, and we found that it leads to higher accuracies in PINN's predictions.

The PINN optimises a loss function consisting of various terms:
\begin{equation}
\begin{split}
\mathcal{L} & = \mathcal{L}_\text{eq}+  \mathcal{L}_\text{IC}+ \mathcal{L}_\text{BC}+ \mathcal{L}_\text{data}\\
& = \frac{1}{N_\text{eq}} \sum_{j=1}^{N_\text{eq}} (f_V({x_j},t_j)^2 + f_W({x_j},t_j)^2) + \frac{1}{N_\text{IC}} \sum_{l=1}^{N_\text{IC}} (V({x_l},t_0) - V_0) ^2  \\
& + \frac{1}{N_\text{BC}} \sum_{k=1}^{N_\text{BC}} (\frac{\partial V}{\partial \vec{n}}(x_k,t_k))^2  + \frac{1}{N_\text{data}} \sum_{i=1}^{N_\text{data}} (V({x_i},t_i)- {V_\text{GT}}_i)^2.\\
\end{split}
    \label{eq:loss} 
\end{equation}

$\mathcal{L}_\text{eq}$ quantifies the agreement with the EP models: AP (Supplementary Eq.~1a \& 1b) or FK (Supplementary Eq.~2a-2c). Here, we use the AP model as an example, denoted by $f_V$ and $f_W$. In the case of the FK model, $\mathcal{L}_{eq}= \frac{1}{N_\text{eq}} \sum_{j=1}^{N_\text{eq}} (f_u({x_j},t_j)^2 + f_v({x_j},t_j)^2 +f_w({x_j},t_j)^2)$.
 $\mathcal{L}_\text{IC}$ is the agreement with the initial conditions of the AP or FK equations. $\mathcal{L}_\text{BC}$ represents the agreement with the no-flux boundary condition for $V$, or $u$ in FK model. $\mathcal{L}_\text{data}$ is the agreement with the training experimental ground truth data, denoted by $V_\text{GT}$. We only use data for the transmembrane potential $V$, or $u$ in FK model, for the $\mathcal{L}_\text{data}$ term, since the other model variables are not experimentally measurable. 
 
 All of these agreements are quantified as mean squared errors and evaluated on the number of collocation points, $N_\text{eq}$, $N_\text{IC}$, $N_\text{BC}$, and $N_\text{data}$, detailed in Supplementary Table 3. Each term is given an equal weight, as preliminary tests showed that this led to the best outcomes.

For EP parameter estimation (Aims 3a \& 3b), the unknown parameter of interest is included as an additional trainable parameter that the network will optimise as it trains with the combined loss in \eqref{eq:loss}. The initial guess for each parameter is randomly taken from a uniform distribution centred at the ground truth value.

As the primary metric to assess PINNs' performance, we calculate the root mean square error (RMSE) between the ground truth and predictions for membrane potential $V$ (or $u$ in FK model) across all test points, which is 80\% of all simulated data:
\begin{equation}
    \text{RMSE}=\sqrt{\frac{1}{N_\text{test}}\sum_{i=1}^{N_\text{test}}(V(x_i,t_i)-V_\text{GT}(x_i,t_i))^2}.
    \label{eq:RMSE}
\end{equation}
We train the model using 1 RTX6000 GPU and 4 CPUs. Further details about the model hyperparameters can be found in Supplementary Table 3.

\begin{figure}[h!]
    \centering
\includegraphics[width=0.9\textwidth]{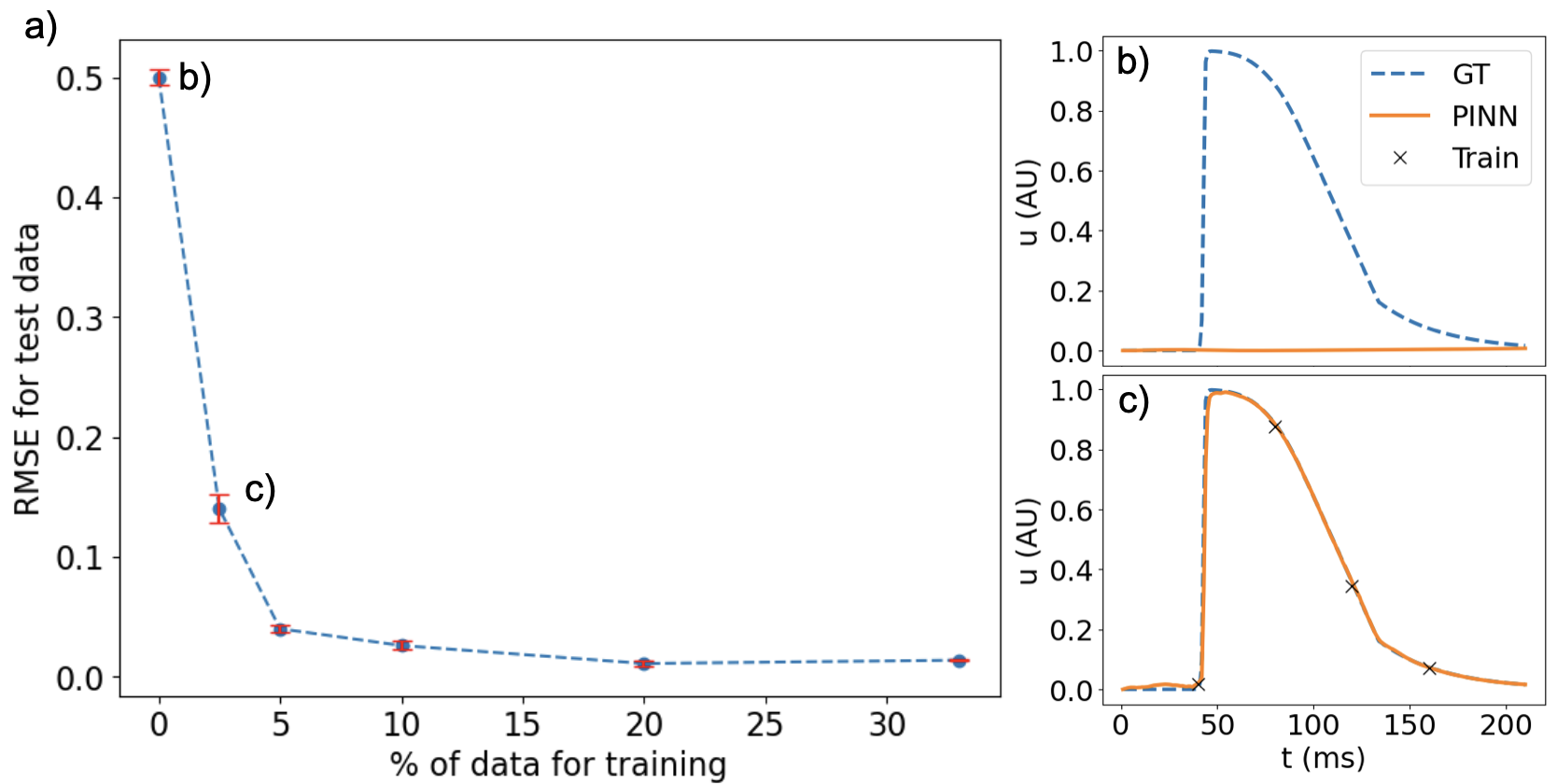}
    \caption{The effects of training data size on PINNs' performance in forward mode. These experiments were conducted in the 2D planar wave scenario with the FK model (Aim 2a), where the total number of data points is $2.1 \times 10^6$. (a) The RMSE for test data, calculated as in \eqref{eq:RMSE}, is plotted as a function of the percentage of training data given. (b) \& (c) are examples of transmembrane potential $u$ across time sampled at a fixed location, in the case of zero and $2.5\%$ training data, respectively.}
    \label{fig:train_size}
\end{figure}

\textbf{Choice of train-test split ratio}. In forward mode, PINNs should technically be able to compute forward solutions without any training data $\mathcal{L_\text{data}}$, but only using $\mathcal{L_\text{eq}}$, $\mathcal{L_\text{IC}}$, and $\mathcal{L_\text{BC}}$, as this specifies a unique solution to the system of equations. It is with the inverse parameter estimation that data are necessary. To test this, we look at the effects of training data size on PINNs' performance in forward experiments, measured by RMSE on the test data. With zero data sample provided, PINNs struggled to learn as shown in Fig.~\ref{fig:train_size}b, with a poor RMSE of $0.50\pm0.01$. As the training data size increases, the RMSE first drops quickly and then plateaus (Fig.~\ref{fig:train_size}a). Based on these results, we keep $\mathcal{L_\text{data}}$ in the forward experiments and choose to use 20\% of the data for training, as it is a balance between performance and size.




\section{Results}
\subsection{Forward Solution of EP models}
\begin{figure}[htp]
\centering
\includegraphics[width=1.0\textwidth]{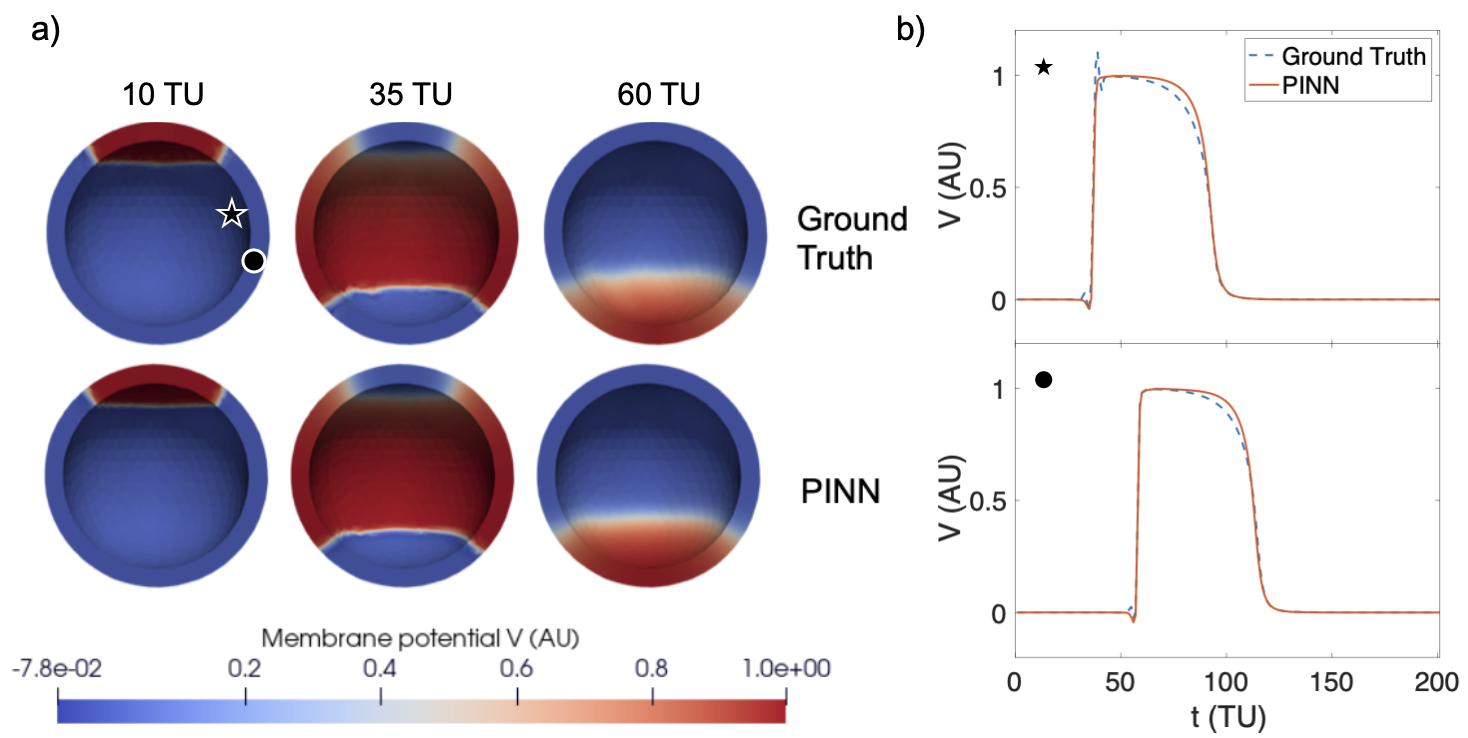}
\caption{A centrifugal wave on a 3D spherical shell with the AP model. (a) A visual comparison between the ground truth data from FEM solver and PINN's predictions, showing transmembrane potential $V$ at various time points. The full movie can be found in Supplementary Video 1. (b) The time evolution of $V$ sampled at two different locations, indicated by the star and the circle signs.}
\label{fig:planar_sphere}
\end{figure}
\subsubsection{PINNs with AP model in 3D}
For a centrifugal wave on a spherical surface modelling sinus rhythm, PINNs accurately reproduced the action potential propagation with an RMSE of $(2.3\pm0.3)\times10^{-2}$ 
on test data. A visual comparison between the ground truth data and PINNs' predictions is shown in Fig.~\ref{fig:planar_sphere}. PINNs reproduces the action potentials well, with some discrepancies with the FEM solver at the wavefront and waveback (Supplementary Video 1.1).

\begin{figure}[htb]
\centering
\includegraphics[width=1.0\textwidth]{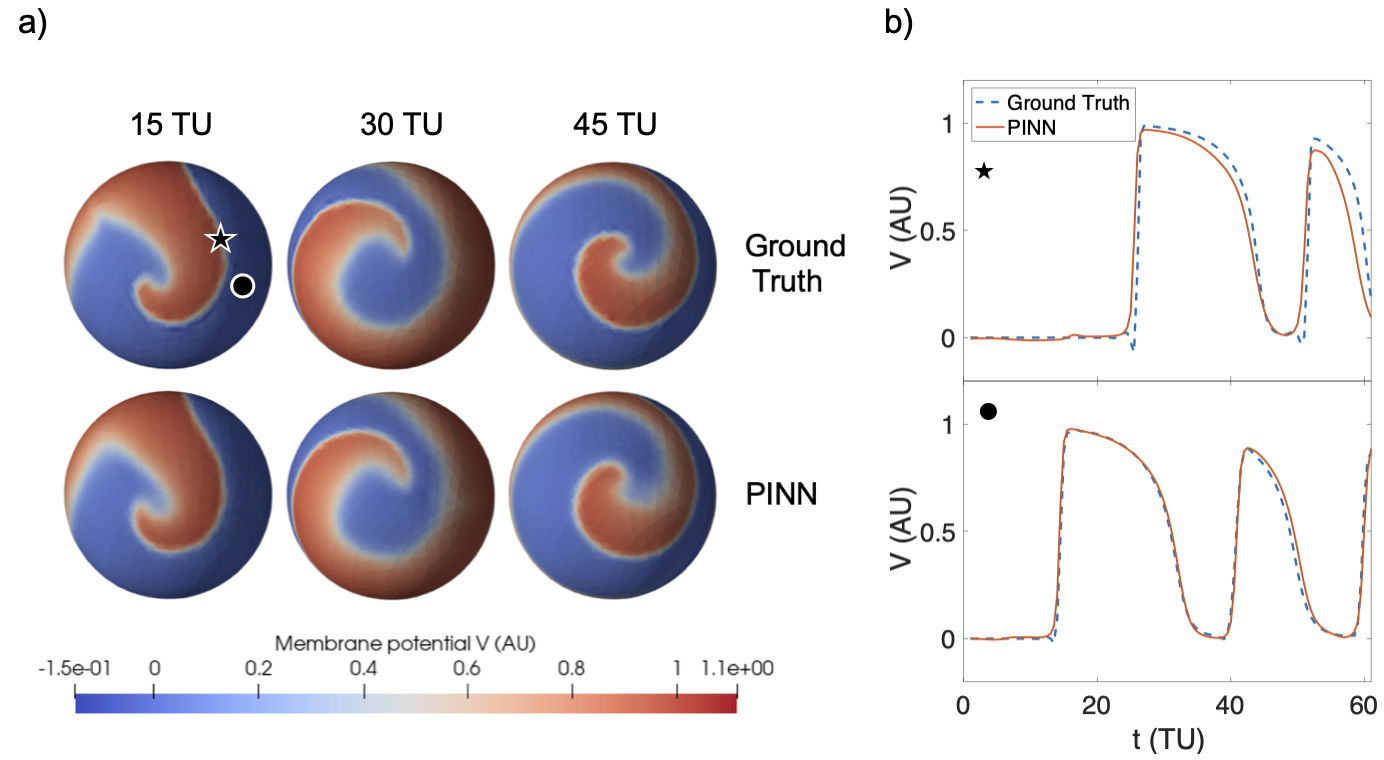}
\caption{A spiral wave on a 3D spherical shell with the AP model. (a) A visual comparison between ground truth data for $V$ and PINN's prediction at various time points (full movie in Supplementary Video 2). (b) The time evolution of $V$ sampled at two different locations.}
\label{fig:spiral_sphere}
\end{figure}

For a spiral wave on a spherical surface, PINNs achieved an RMSE of $(3.7 \pm 0.5) \times10^{-2}$ 
on test data, slightly higher than the planar wave case but still $<4\%$ of the peak $V$ value. Fig.~\ref{fig:spiral_sphere} gives the visualisation of the results. The biggest differences between the FEM ground truth data and PINN's prediction were found at the boundaries and tips of the spiral waves (Supplementary Video 2.1).

\subsubsection{PINNs with AP model in fibrillatory conditions (2D)}

\begin{figure}[htb]
\centering
\includegraphics[width=1.0\textwidth]{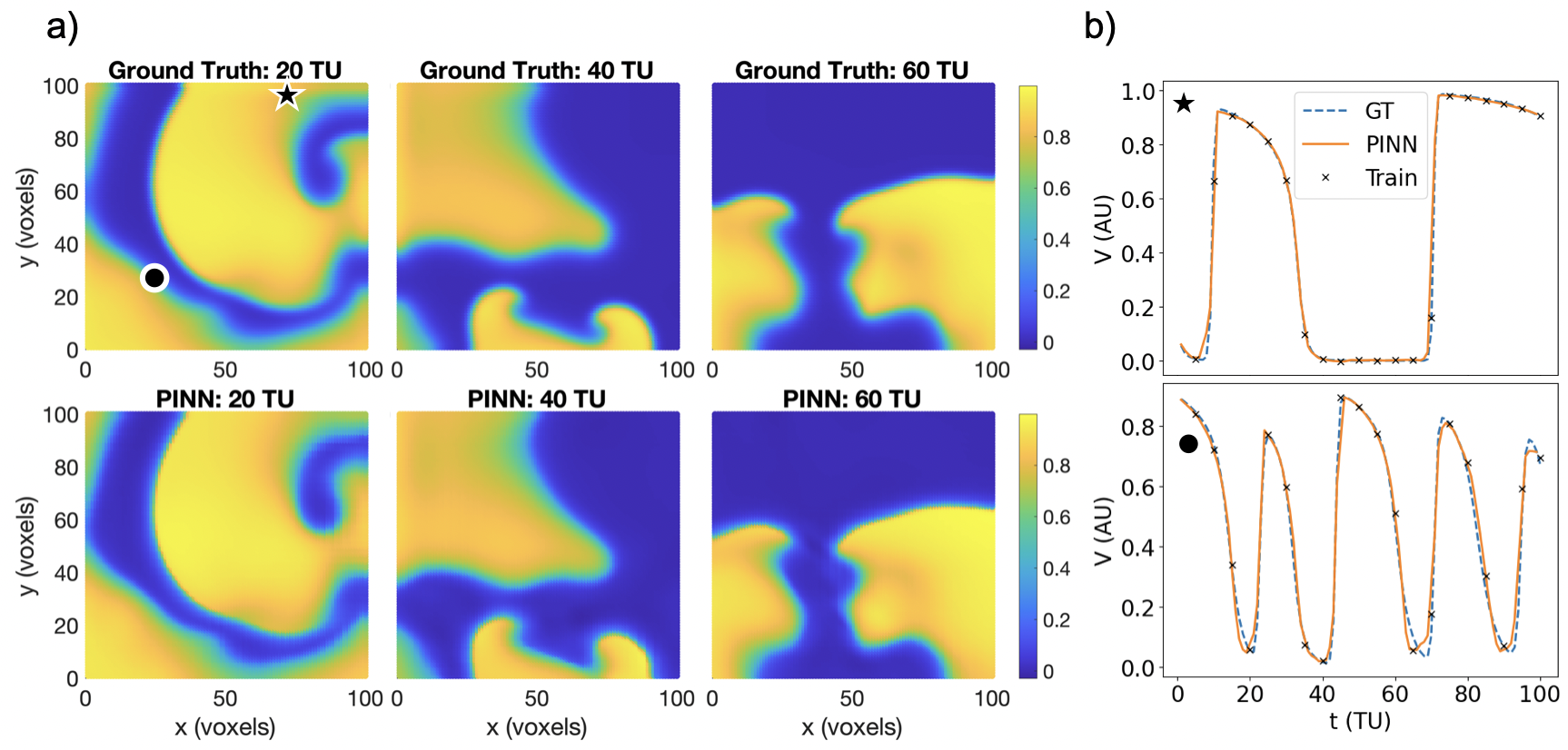}
\caption{Spiral wave break-up dynamics on a 2D rectangle with the AP model. (a) Comparison between the ground truth data generated by finite differences method and PINN's predictions for $V$ at various time points (Supplementary video 3). (b) $V(t)$ sampled at two different locations.}
\label{fig:breakup_2D}
\end{figure}

In the 2D spiral wave break-up regime, PINNs achieved an RMSE of $(5.1\pm2.6) \times 10^{-2}$ 
on test data (Fig.~\ref{fig:breakup_2D}). The performance of PINNs in this scenario was particularly sensitive to the network initialisation, leading to a larger variance in the RMSEs.

\subsubsection{PINNs with FK model in 2D}
\begin{figure}[h!]
    \centering
\includegraphics[width=1.0\textwidth]{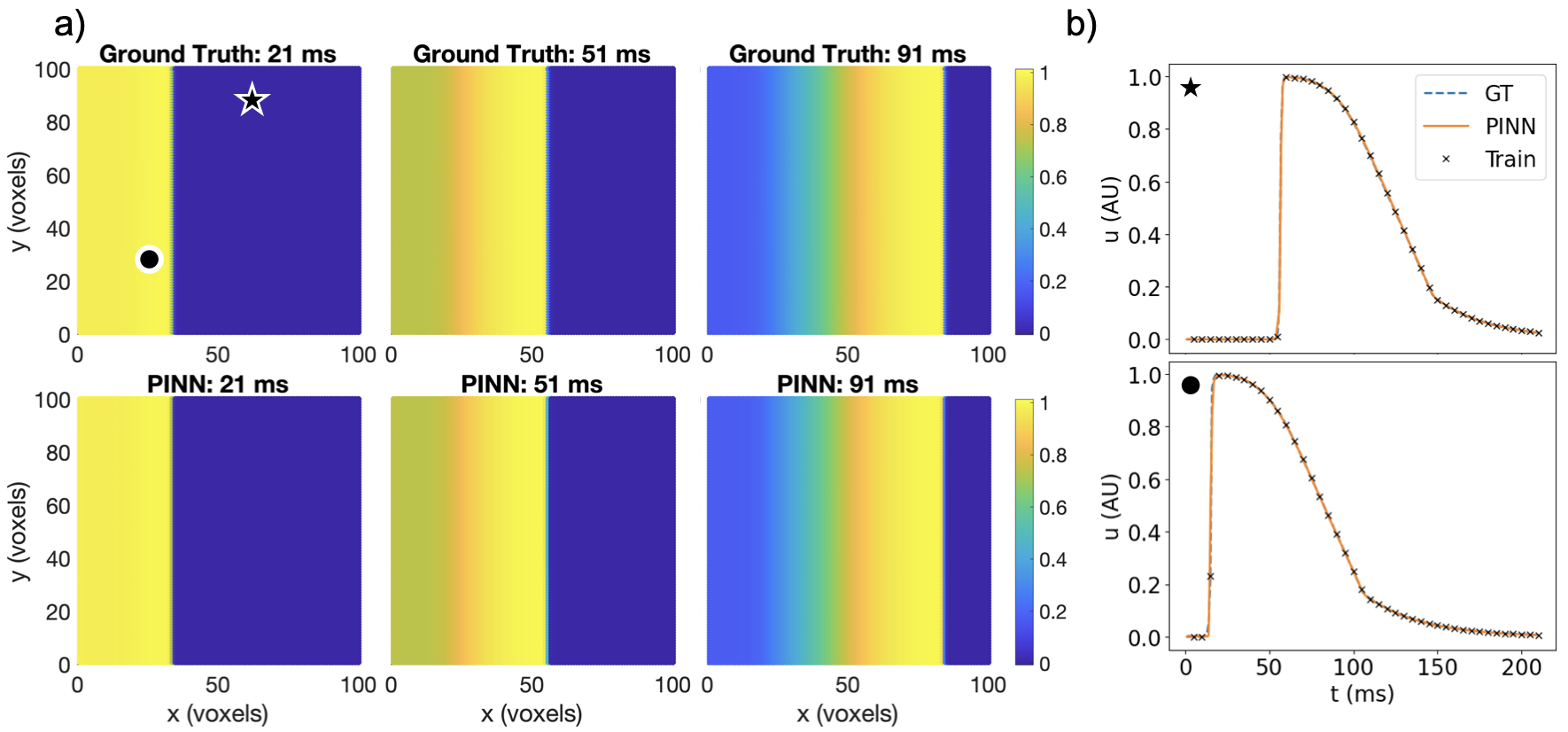}
    \caption{A planar wave on a 2D rectangle with the FK model. (a) Comparison between the ground truth data and PINN's predictions for $u$ at various time points (Supplementary Video 4). (b) $u(t)$ sampled at two different locations.}
    \label{fig:2DFKplanar}
\end{figure}

\begin{figure}[h!]
    \centering
\includegraphics[width=1.0\textwidth]{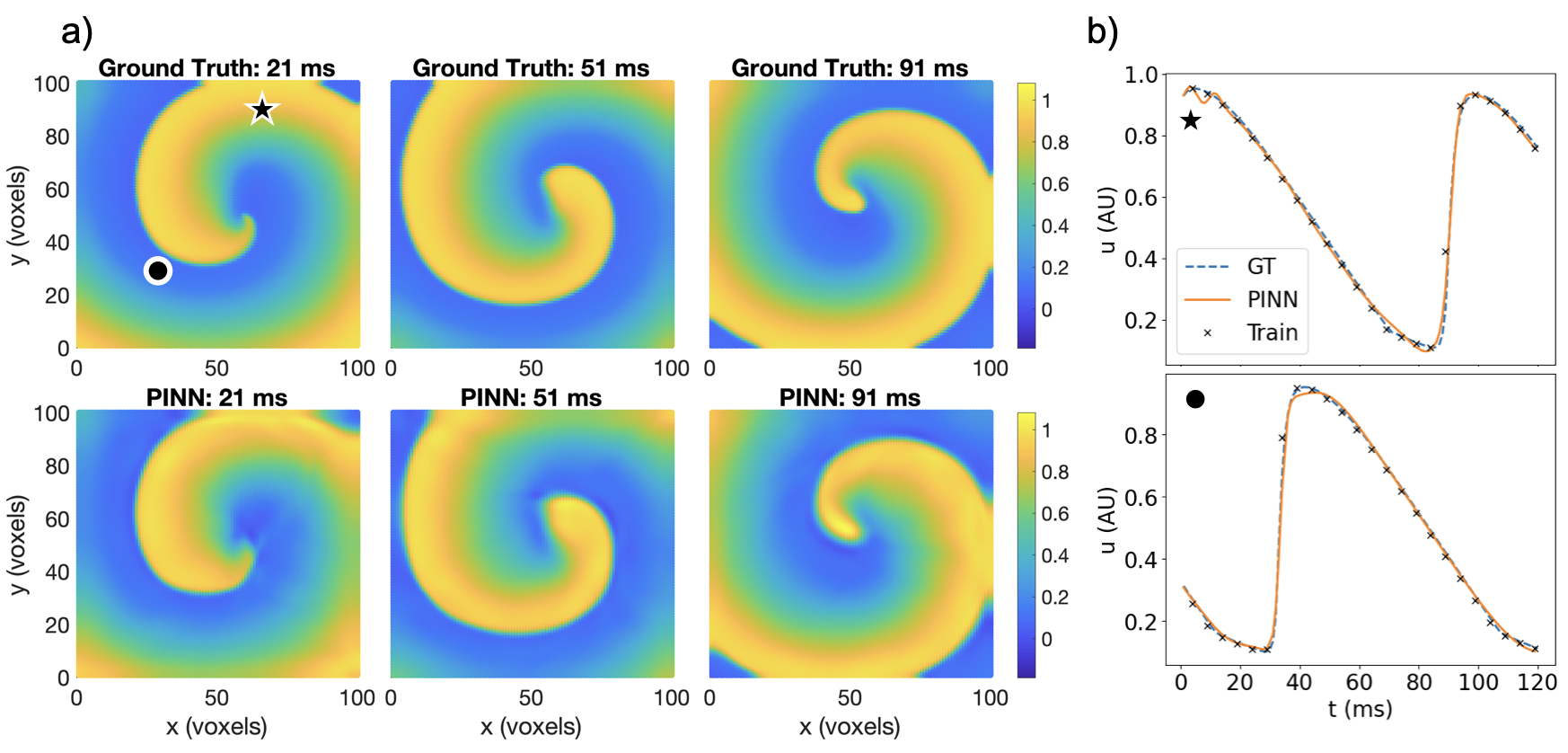}
    \caption{A spiral wave on a 2D rectangle with the FK model. (a) Comparison between the ground truth and PINN's predictions for $u$ at various time points (Supplementary Video 5). (b) $u(t)$ sampled at two different locations.}
    \label{fig:2DFKspiral}
\end{figure}

Moving to the more complex FK model, for planar waves on a rectangle, PINNs achieved an excellent RMSE of $(1.1 \pm0.2 )\times 10^{-2} $. For the spiral wave, the RMSE increased to ($2.9 \pm 0.5) \times 10^{-2}$, comparable to the accuracy of PINNs using the AP model. The visualisations of these results are shown in Fig.~\ref{fig:2DFKplanar} and~\ref{fig:2DFKspiral}. 

\subsection{Inverse estimation of EP parameters}
\begin{figure}[h!]
    \centering
\includegraphics[width=0.6\textwidth]{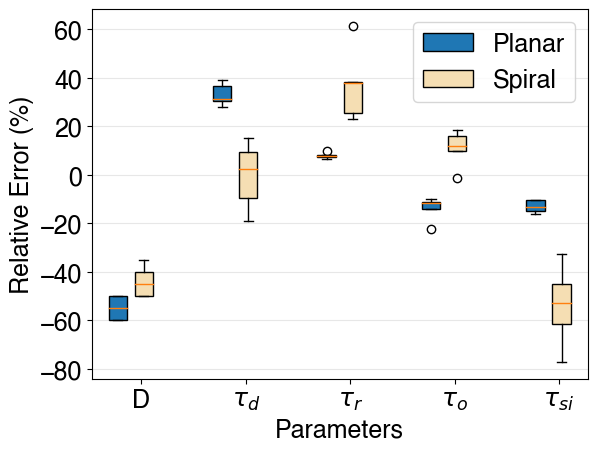}
    \caption{Relative errors in the parameter estimates for the FK model in 2D, for both planar and spiral wave regimes.}
    \label{fig:RE_para_est}
\end{figure}

For the FK model in 2D, we used PINNs in inverse mode to estimate the isotropic diffusion coefficient $D$, as well as parameters $\tau_d$, $\tau_r$, $\tau_o$, and $\tau_{si}$ in Supplementary Eq.~3a-3c
, which are approximately inversely related to ion channel conductances. The RMSE for $u$ on test data in inverse mode was $(9.9\pm0.9)\times10^{-3}$ for planar waves and $(3.1\pm0.4)\times 10^{-2}$ for spiral waves, similar to the forward mode RMSEs. The parameter estimation results are summarised in Fig.~\ref{fig:RE_para_est}. 

PINNs were able to estimate the FK model parameters accurately in both 2D scenarios. For the Ca\textsuperscript{2+} and K\textsuperscript{+} channel conductance ($\tau_{si}$ and $\tau_r$), PINNs achieved relative errors (REs) $< 15\%$ for planar waves, and the REs were larger in the spiral wave regime. For $\tau_o$, the REs were similar in both regimes, both within $20\%$. For $\tau_d$, related to the Na\textsuperscript{+} channel, the average performance was actually better in the spiral wave regime ($<0.5\%$). The estimation of the diffusion coefficient $D$ in both regimes and $\tau_{si}$ for spiral wave were noticeably poorer ($> 40\%$) than all the other parameters ($<40\%$).

\section{Discussion}
We present a PINN framework capable of accurately reconstructing cardiac action potentials in various geometries and dynamical regimes, including 3D spherical surfaces and fibrillatory conditions in 2D. This was done without any data for the latent variables but only the transmembrane potentials, hinting at the possibility of using PINNs for clinical cardiac digital twins~\cite{niederer2019computational}.

In general, PINNs predicted action potential propagation at a higher accuracy in sinus rhythm conditions than for spiral waves, in both 2D and 3D geometries. In sinus rhythm, PINNs achieved better performance in 2D than 3D, although a more complex model was used in 2D. Compared to Herrero Martin \textit{et al.}'s work~\cite{HerreroMartin2022EP-PINNs:Networks}, which reported RMSE $< 3.0 \times 10^{-2}$ using the AP model in the same 2D rectangular geometry, our work with the more complex FK model actually achieved a slightly lower mean RMSE, $(1.1 \pm0.2 )\times 10^{-2} $, and lower upper bound RMSE $<1.3\times10^{-2}$ throughout. 
For spiral waves in 2D, our RMSE 
was slightly higher than Herrero Martin \textit{et al.}'s. 
These results show that PINNs' performance depends on both the EP model complexity and the dynamical regime of the solution. Among all considered scenarios, PINNs performed the worst in the 2D fibrillatory condition, which is characterised by complex waveforms involving rapid non-periodic changes in transmembrane potential and latent variables.

PINN was additionally able to estimate, with varying degrees of accuracy (mean $-0.09\pm 0.33$), parameters related to ion channel conductances in the FK model in 2D geometries. This could have applications in understanding the effects of disease-induced remodelling or anti-arrhythmic drugs on different ion channels~\cite{chiu2024characterisation} in geometries more closely resembling the heart's. 
From our experiments, we note that a lower RMSE \eqref{eq:RMSE} does not necessarily correlate with a lower error in parameter estimation, which makes the refinement of parameter estimation more complicated. PINNs' convergence is also particularly difficult in inverse problems~\cite{wang2023expert}. 




The results were obtained in simple and homogeneous media, which might limit the application of our method in a clinical scenario. In the future, we plan to train PINNs on realistic heart shapes and in anisotropic conditions. We also aim to extend beyond \textit{in silico} data, training with experimental optical mapping data from cardiac tissue~\cite{lebert2021rotor} or electrogram signals to bring this framework closer to clinical applications. PINNs are unique in their ability to simultaneously solve model equations and perform parameter estimation, which makes them great candidate for cardiac digital twins~\cite{niederer2019computational}. Our study establishes PINNs' potential for these applications by showing that they can be successfully deployed in 3D and fibrillatory conditions.

\subsection*{Acknowledgments}
This work was supported by the NIHR Imperial Biomedical
Research Centre (BRC), the British Heart Foundation Centre of Research Excellence at Imperial College London (RE/18/4/34215). The neural network training was supported by the Imperial College Research Computing Service (DOI: 10.14469/hpc/2232). 

\bibliographystyle{splncs04} 
\bibliography{references,refs}

\end{document}


\title{Physics-Informed Neural Networks can accurately model cardiac electrophysiology in 3D geometries and fibrillatory conditions: Supplementary Materials}
\titlerunning{PINNs for Cardiac Electrophysiology: Supplementary Materials}
\author{Ching-En Chiu\inst{1,2} \and Aditi Roy\inst{3} \and Sarah Cechnicka\inst{4} \and Ashvin Gupta\inst{5} \and Arieh Levy Pinto\inst{5} \and Christoforos Galazis\inst{4} \and Kim Christensen\inst{6} \and Danilo Mandic\inst{1} \and Marta Varela\inst{2}}
 \authorrunning{Chiu \textit{et al}.}
%

\institute{Department of Electrical and Electronic Engineering, Imperial College London, UK \and 
National Heart \& Lung Institute, Imperial College London, UK \and
Department of Biomedical Engineering, Kings College London, UK \and
Department of Computing, Imperial College London, UK \and
Department of Bioengineering, Imperial College London, UK \and
Department of Physics, Imperial College London, UK 
\\
\email{marta.varela@imperial.ac.uk}}
\maketitle

\section{Supplementary Equations}
\subsection{Aliev-Panfilov Model}
The Aliev-Panfilov model equations are as follows:
\begin{subequations}
\begin{equation}\label{eq:AlievPanfilov1}
\frac{\partial V}{\partial t} = \nabla \cdot (D \nabla V) - kV (V - a) (V - 1) - VW
\end{equation}
\begin{equation}\label{eq:AlievPanfilov2}
\frac{dW}{dt} = \left( \epsilon + \frac{\mu_1 W}{V + \mu_2} \right) \left( -W - kV (V - b - 1) \right).
\end{equation}
\end{subequations}

The nonlinear term $-kV(V-a)(V-1)-VW$ describes the changes in transmembrane potential $V$ due to different ionic fluxes. $a$ is a parameter related to the threshold potential, above which an AP can be initiated. $W$ is a non-observable recovery variable, which is related to the restitution and enforces refractoriness properties. Parameter $b$ controls the refractoriness. The variables are dimensionless. 
\subsection{Fenton-Karma Model}
The Fenton-Karma model equations are as follows:
\begin{subequations}
\begin{equation}
\label{FK1} \partial_t u = \nabla \cdot (D\nabla u) - J_\text{fi}(u;v) - J_\text{so}(u) - J_\text{si}(u;w),
\end{equation}
\vspace*{-0.25cm}
\begin{equation}
\partial_t v = \Theta(u_c - u)(1 - v)/\tau_v^{-}(u) - \Theta(u - u_c)v/\tau_v^{+}, \label{FK2}
\end{equation}
\vspace*{-0.25cm}
\begin{equation}
\partial_t w = \Theta(u_c - u)(1 - w)/\tau_w^{-} - \Theta(u - u_c)w/\tau_w^{+}, \label{FK3}
\end{equation}
\end{subequations}where the three scaled currents are given by
\begin{subequations}
\begin{equation}\label{eq:Jfi}
J_\text{fi}(u,v) = -\frac{v}{\tau_d} \Theta(u - u_c)(1 - u)(u - u_c),
\end{equation}
\vspace*{-0.25cm}
\begin{equation}\label{eq:Jso}
 J_\text{so}(u) = \frac{u}{\tau_o} \Theta(u_c - u) + \frac{1}{\tau_r} \Theta(u - u_c),
 \end{equation}
 \vspace*{-0.25cm}
 \begin{equation}\label{eq:Jsi}
 J_\text{si}(u;w) = -\frac{w}{2\tau_\text{si}} (1 + \tanh[k(u - u_c^\text{si})]).
 \end{equation}
 \end{subequations}

The three variables are: $u$, the dimensionless membrane potential scaled to $[0,1]$ (corresponding to $V$ in the AP model), and $v$ and $w$, latent gate variables for $J_\text{fi}$ and $J_\text{si}$, respectively. $\Theta(x)$ is the Heaviside function. The $\tau$ parameters are various time constants. As shown in Eq.\eqref{eq:Jfi}-\eqref{eq:Jsi}, $\tau_d$, $\tau_o$, $\tau_r$, and $\tau_{si}$ are approximately inversely related to the conductance of different ion channels. 

We assumed homogeneous and isotropic diffusion of the membrane potential throughout. Therefore, the $\nabla \cdot (D \nabla V)$ term in both models reduces to $D \nabla^2 V$, where $D$ is a scalar diffusion coefficient as opposed to a tensor.

\section{Supplementary Tables}
\begin{table}[h!]
    \centering
    \begin{tabular}{| m{3cm} | m{1cm} | m{1cm} | m{1cm} | m{1cm} | m{0.7cm} | m{0.7cm} | m{0.7cm} | m{1.56cm} |}
        \hline
        \textbf{Aliev-Panfilov} & D & a & b & $\epsilon$ & $\mu_1$ & $\mu_2$ & k & Duration\\
        \hline
        2D spiral break-up & 0.15 & 0.05 & 0.05 & 0.002 & 0.2 & 0.3 & 8.0 & 400 TU, using the final 100TU \\
        \hline
        3D centrifugal waves & 0.1 & 0.01 & 0.15 & 0.002 & 0.2 & 0.3 & 8.0 & 201 TU \\
        \hline
        3D spiral waves & 0.1 & 0.01 & 0.15 & 0.002 & 0.2 & 0.3 & 8.0 & 121 TU \\
        \hline
    \end{tabular}
    \caption{Parameter values used for generating different dynamical regimes using the Aliev-Panfilov model. For 2D spiral break-up, the values were taken from \cite{lebert2023reconstruction}. For the 3D scenarios, the values were taken from \cite{goktepe2010electromechanics} and  \cite{nash2004electromechanical}. }
    \label{table:param_value_AP}
\end{table}

\begin{table}[h!]
\centering
\begin{tabular}{| m{3cm} | m{1cm} | m{1cm} | m{1cm} | m{1cm} | m{1cm} | m{1cm} | m{1cm} | m{1.5cm} |}
\hline
\textbf{Fenton-Karma} & D & $\tau_d$ & k & $\tau_r$ & $\tau_{si}$ & $\tau_o$ & $\tau_{v+}$ & Duration \\ \hline
\multirow{2}{*}{2D planar waves} 
& 0.02 & 0.125 & 10 & 70 & 114 & 32.5 & 5.8  & 210 TU\\ \cline{2-8}
 & $\tau_{v1}^-$ & $\tau_{v2}^-$ & $\tau_{w}^+$ & $\tau_{w}^-$ & $u_c$ & $u_v$ & $u_c^{si}$ \\ \cline{2-8} \hline
 2D spiral waves & 82.5 & 60 & 300 & 400 & 0.16 & 0.04 & 0.85 & 120 TU\\ \hline
\end{tabular}
\caption{Parameter values used for generating both planar and spiral waves using the Fenton-Karma model in 2D. The values were taken from the FK-CAF model in \cite{goodman2005membrane}.}
\end{table}

\begin{table}[htb]
    \centering
    \renewcommand{\arraystretch}{1.5}
    \begin{tabular}{|m{1.4cm}|m{2cm}|m{1.6cm}|m{1.5cm}|m{1.5cm}|m{1.4cm}|m{1.5cm}|m{1.5cm}|m{1.6cm}|}
        \hline
        & \textbf{Geometry/ dynamics} & \textbf{Ground truth duration} & \textbf{Training time} & \textbf{Learning rate} & \textbf{Epochs for Adam} & \textbf{training points for $\mathcal{L}_{eq}$} & \textbf{training points for $\mathcal{L}_{BC}$} & \textbf{Network size} \\
        \hline
        \textbf{Aim 1a}  & 3D AP centrifugal & 201 TU& $\sim$7hr& 0.0005 & 165,000 & 40000 & 4000 & 4 layers x 32 neurons \\
        \hline
        \textbf{Aim 1b} & 3D AP spiral & 121 TU& $\sim$5hr & 0.0005 & 165,000 & 40000 & 4000 & 4 layers x 32 neurons \\
        \hline
        \textbf{Aim 1c} & 2D AP break-up &  100 TU & $\sim$9hr & 0.0005 & 150,000 & 30000 & 1000 & 5 layers x 60 neurons \\
        \hline
        \textbf{Aim 2a} 
        & 2D FK planar & 210 ms & $\sim$8hr & 0.0005 & 150,000 & 30000 & 1000 & 4 layers x 32 neurons \\
        \hline
        \textbf{Aim 2b} & 2D FK spiral & 120 ms& $\sim$5hr & 0.0005 & 150,000 & 30000 & 1000 & 4 layers x 32 neurons \\
        \hline
        \textbf{Aim 2c} & 2D FK planar (inverse) & 210 ms & $\sim$8hr & 0.0005 & 150,000 & 30000 & 1000 & 4 layers x 32 neurons \\
        \hline
        \textbf{Aim 2d} & 2D FK spiral (inverse) & 120 ms& $\sim$5hr & 0.0005 & 150,000 & 30000 & 1000 & 4 layers x 32 neurons \\
        \hline
    \end{tabular}
    \caption{Details of training hyperparameters for each geometry and dynamical regime.}
    \label{tab:training_params}
\end{table}

\section{Supplementary Videos}
The attached supplementary videos show a comparison between the ground truth data and the PINN's predictions.


\begin{itemize}
    \item Video 1: planar wave on a 3D spherical shell, Aliev-Panfilov model.
    \item Video 1.1: the absolute errors between ground truths and PINN's predictions. Planar wave on a 3D spherical shell, Aliev-Panfilov model.
    \item Video 2: spiral wave on a 3D spherical shell, Aliev-Panfilov model.
    \item Video 2.1: the absolute errors between ground truths and PINN's predictions. Spiral wave on a 3D spherical shell, Aliev-Panfilov model.
    \item Video 3: spiral wave break ups on a 2D square, Aliev-Panfilov model.
    \item Video 4: planar wave on a 2D square, Fenton-Karma model.
    \item Video 5: spiral wave on a 2D square, Fenton-Karma model.
\end{itemize}

\bibliographystyle{splncs04} 
\bibliography{references,refs}